# Title

Toward parallel intelligence: an interdisciplinary solution for complex systems

# Authors

Yong Zhao[1†], Zhengqiu Zhu[1†], Bin Chen[1,2†], Sihang Qiu[1†*], Jincai Huang[1,2*], Xin Lu[1,3*], Weiyi Yang[1], Chuan Ai[1], Kuihua Huang[1,2], Cheng He[3,4], Yucheng Jin[5], Zhong Liu[1,2], Fei-Yue Wang[6]

# Affiliations

[1] College of Systems Engineering, National University of Defense Technology, Changsha, 410073, China.

[2] Hunan Institute of Advanced Technology, Changsha, 410073, China.

[3] Shanghai Institute of Infectious Disease and Biosecurity, Shanghai, 200032, China.

[4] Institute of Epidemiology, Helmholtz Zentrum München – German Research Center for Environmental Health (GmbH), Neuherberg, Germany.

[5] Department of Computer Science, Hong Kong Baptist University, Hong Kong, 999077, China.

[6] State Key Laboratory for Management and Control of Complex Systems, Institute of Automation, Chinese Academy of Sciences, Beijing 100190, China.

[†] These authors contributed equally to this work.

[*] Corresponding authors: qiusihang11@nudt.edu.cn (S.-H.Q.); huangjincai@nudt.edu.cn (J.-C.H.); xin.lu.lab@outlook.com (X.L.).

# PUBLIC SUMMARY

- Interdisciplinary research stimulates the development of parallel systems method.
- ACP approach is a technical foundation to address challenges in complex systems.
- Parallel intelligence is objective for a better understanding of complex systems.
- Various parallel technologies and applications are major accomplishments.

# GRAPHIC ABSTRACT

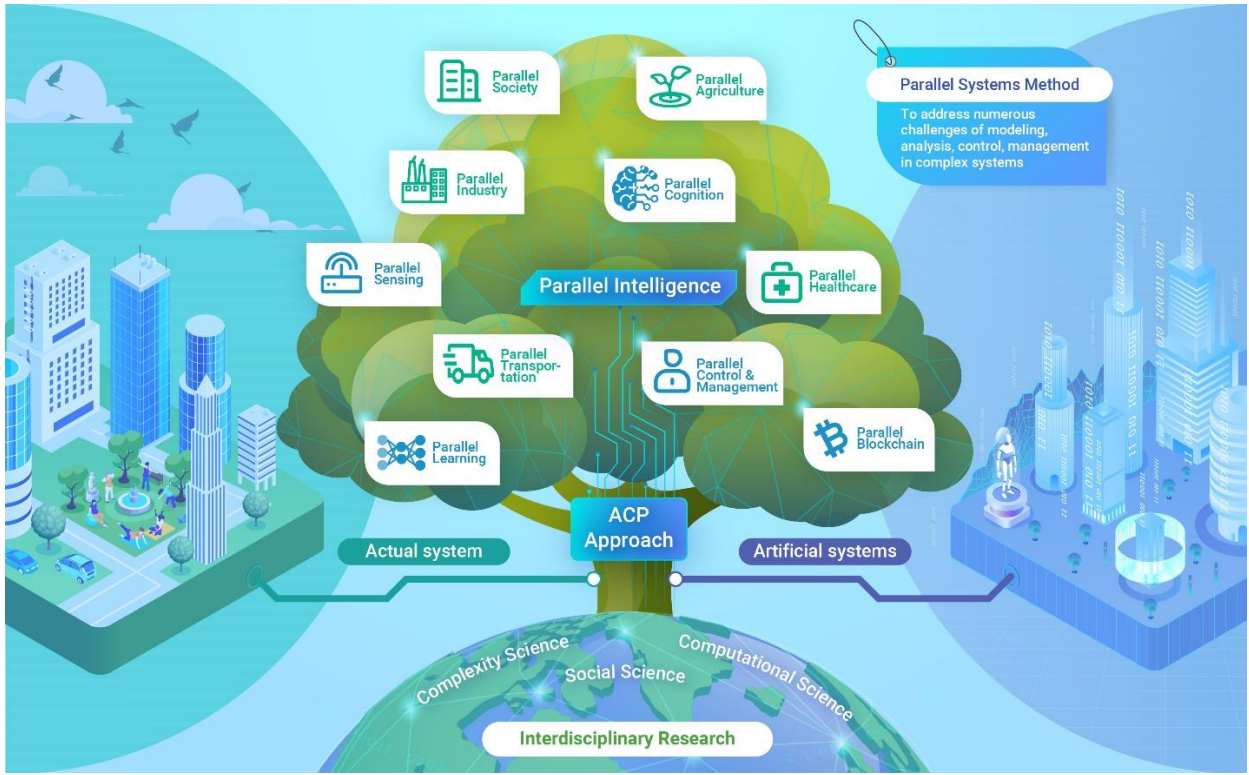

## ABSTRACT

The growing complexity of real-world systems necessitates interdisciplinary solutions to confront myriad challenges in modeling, analysis, management, and control. To meet these demands, the parallel systems method rooted in Artificial systems, Computational experiments, and Parallel execution (ACP) approach has been developed. The method cultivates a cycle, termed parallel intelligence, which iteratively creates data, acquires knowledge, and refines the actual system. Over the past two decades, the parallel systems method has continuously woven advanced knowledge and technologies from various disciplines, offering versatile interdisciplinary solutions for complex systems across diverse fields. This review explores the origins and fundamental concepts of the parallel systems method, showcasing its accomplishments as a diverse array of parallel technologies and applications, while also prognosticating potential challenges. We posit that this method will considerably augment sustainable development while enhancing interdisciplinary communication and cooperation.

## 1. INTRODUCTION

Stephen Hawking's prediction that the 21st century would be a "century of complexity"[1] has been substantiated over the last two decades. Real-world complex systems, such as those in engineering and social systems, manifest remarkable complexities. These include non-linearity, randomness, networked coupling, collective dynamics, hierarchy, and emergence, all of which contribute to substantial uncertainty and unpredictability in the managing and controlling these systems. This complexity can be partly attributed to advances in Information and Communications Technology (ICT), enabling the Internet of Everything.[2] Such extensive connectivity promotes the growth of complex systems with larger scales, more intricate structures, and nonlinear interactions between elements of varying magnitudes across different domains. Examples include social media and other cyber applications, physical space's Internet of Things (IoT) nodes, and heterogeneous individuals in social space.[3] The cross-domain interactions enable the integration of cyber, physical, and social spaces in complex systems,[4] thereby unifying the dual complexity of social and engineering (or technical) aspects into a single system. This heightens the challenges in complex systems concerning modeling, analysis, management, and control. As a result, tackling these issues using the knowledge and technology of a single discipline is a daunting task. To address the increasing complexity of systems, interdisciplinary research has become essential, combining knowledge and technologies from multiple disciplines to provide a comprehensive analysis and understanding.[5] For instance, various interdisciplinary analytical methods, such as social simulation,[6] data mining,[7] and online experiments[8] have been developed to explore the complexity of systems from diverse perspectives. Additionally, interdisciplinary research paradigms, such as Cyber-Physical-Social Systems (CPSS)[9] or Cyber-Physical Human Systems (CPHS),[10] offer fundamental frameworks for modeling and analyzing systems with the dual complexities of engineering and social aspects. To address the challenges posed by the increasing complexity of real-world systems and drawing inspirations from the burgeoning interdisciplinary research, the Artificial systems, Computational experiments, and Parallel execution (ACP) approach was proposed in 2004.[11] This resulted in the creation of a novel method for complex systems studies—the parallel systems method. Over the past two decades, this method has been continuously evolving, to pursue its objectives of achieving parallel intelligence. The parallel systems method has now emerged as a promising interdisciplinary solution for complex systems, enhancing people's understanding of such systems.[12]

The method of parallel systems discussed in this review is distinct from parallel system or parallel computing in computational science. Unlike these latter concepts, which entail a system capable of executing multiple tasks or instructions simultaneously, the parallel systems method relates to composite systems made up of an actual system and one or more corresponding virtual artificial systems. The concept of digital twins,[13] which is akin to the parallel systems method, is an essential technology for translating actual systems in physical space to digital representations in cyber space. The parallel systems method conveys a broader meaning than digital twins, as it primarily targets complex systems that integrate additional human and social dimensions. In addition to the engineering complexity that stems from large-scale and multi-level systems, social complexity resulting from individual-level human psychology, behavior, and interaction come into play. These factors create challenges in establishing precise models or conducting experimental investigations of complex systems. Recognizing these challenges, the parallel systems method does not aim for complete alignment between the internal mechanisms of artificial and actual systems. Instead, it conducts computational experiments on artificial systems to explore the possible behaviors of the actual system and then implements parallel execution to reconcile any behavioral differences between the two. The parallel execution involves two aspects: first, modifying the artificial systems to emulate the behavior of the actual system; second, steering the actual system toward desired outcomes using insights learned from artificial systems.

The parallel systems method represents an interdisciplinary solution that cohesively integrates theories, technologies, and applications across various disciplines. *1) In terms of theories*, its interdisciplinary character stems from the amalgamation of multiple fields, such as social, computational, and complexity sciences. *2) In terms of technologies*, the method consistently absorbs cutting-edge technologies from diverse disciplines, formulating corresponding parallel technologies to enhance its capacity for dealing with various aspects of complex systems. *3) In terms of applications*, the method leverages parallel technologies and specialized domain knowledge to deploy diverse parallel applications in complex systems across various fields, effectively addressing challenges in modeling, analysis, control, and management.

In this review, we undertake a comprehensive examination of the parallel systems method's origin, highlighting its emergence as a response to the challenges posed by complex systems and the opportunities presented by interdisciplinary research. Moreover, we probe into the details of the method, explicating parallel intelligence as its objectives, the ACP approach as its technical foundation, and advanced parallel technologies. Supported by these technological advancements, we illustrate the implementation and applications of the method in five major fields. Finally, we discuss the challenges and perspectives of the parallel systems method.

## 2. PARADIGM SHIFT FOR UNDERSTANDING COMPLEX SYSTEMS

### 2.1 Increasing complexity of real-world systems

With advancements in scientific theories and technologies, our understanding of the complexities of various real-world systems has deepened significantly. Social systems are quintessential examples of complex systems; the emergence of complex collective behavior and social phenomena at the macro level results from nonlinear interactions among individuals at the micro level. This poses challenges for reductionist analysis and repeatable experimental analyses in social systems. Moreover, with the growing scale and integration of various engineering systems in the real world, the complexity of engineering systems continues to escalate. This complexity poses significant challenges for managing and controlling systems in transportation, energy, manufacturing, and other engineering fields. Nowadays, the integration of social and engineering systems has become an evident trend catalyzed by ICT. This trend is driven by the continuous convergence of cyber, physical, and social spaces,[14] simultaneously augmenting complexity for both social and engineering domains, as depicted in Figure. 1A. Specifically, this is illustrated in two trends.

The first trend pertains to the socialization of engineering systems. In traditional engineering systems, modeling, operation, control, and optimization were considered independently from a technological perspective. However, with increasing concerns regarding safety, sustainability, resilience, eco-friendliness, and human welfare, it is now acknowledged that engineering systems cannot be viewed in isolation from human society.[15] For example, the status of workers in factories can affect production efficiency, cost, and quality;[16] production planning, manufacturing execution, and market demand are interdependent;[15, 17] the state of transportation systems is closely related to social events and human activities.[18–21] Incorporating social and human factors throughout the lifespan of engineering systems is essential for ensuring their sustainability and providing efficient, safe services to society. Achieving this goal requires a more comprehensive understanding of the complex interplay between technical and social factors, along with the development of novel models and methodologies that account for this integration. In general, the socialization of engineering systems represents an important transition towards a more integrated and holistic approach to designing and managing our engineering systems.

Another trend is the engineering of social systems, facilitated by advanced ICT and the integration of ubiquitous sensing devices and IoT technology into human society. This has ushered us into the big data era, where abundant social signals can be leveraged to model, analyze, and operate social systems in a way similar to engineering systems. Concepts such as smart cities,[22] smart societies,[23] and urban computing,[24] have flourished under this trend.

The escalating complexity of systems demands a broader perspective to consider more elements within systems, as well as more powerful tools to understand and analyze them. Consequently, interdisciplinary research has rapidly advanced by merging knowledge and techniques from different disciplines, opening new opportunities for tackling complex system problems. Among the various interdisciplinary research paths, the intersection of computational science and social science offers a range of research methods and paradigms for addressing challenges with both social and engineering complexity.[25]

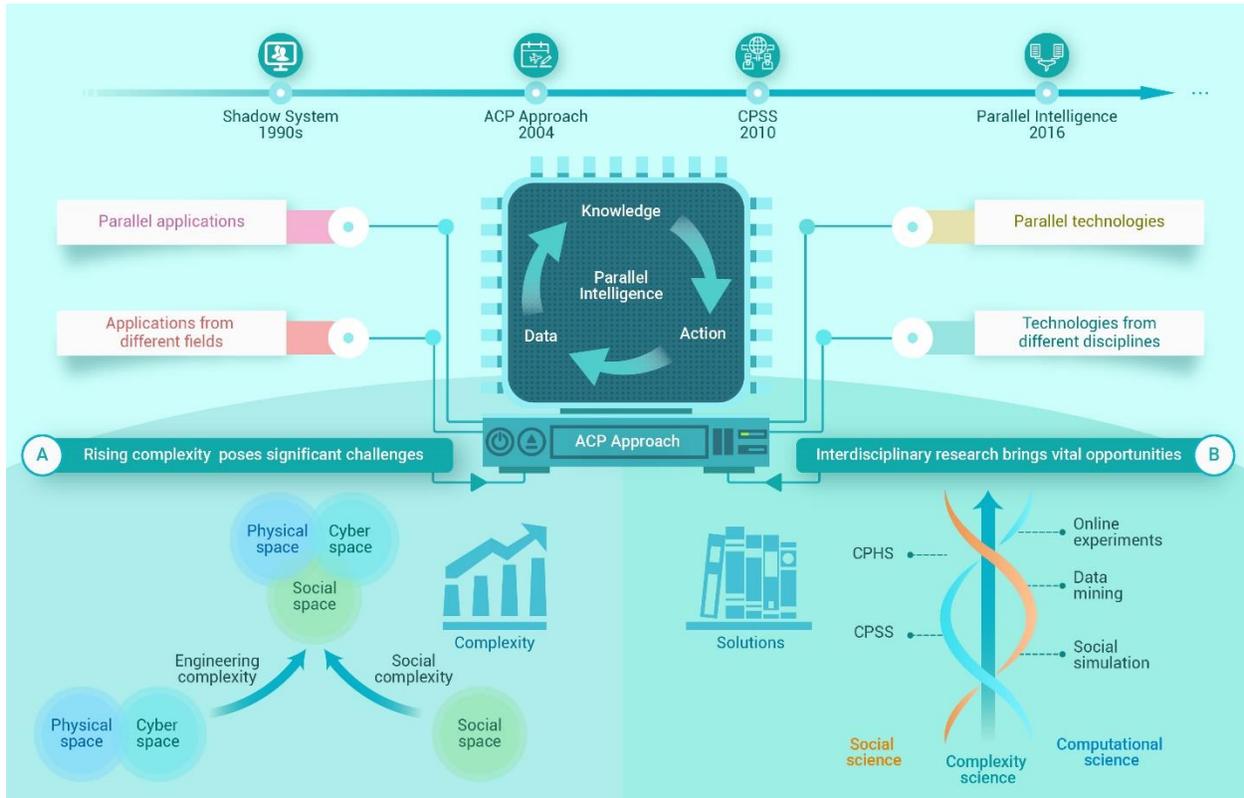

**Figure 1. Advent of the parallel systems method presents a unique opportunity.** (A) Convergence of cyber, physical, and social spaces introduces significant challenges for addressing various problems in complex systems; (B) Interdisciplinary research brings opportunities to address various problems in complex systems. Challenges and opportunities have led to the emergence of the parallel systems method that can absorb advanced technologies from diverse disciplines. Consequently, various parallel technologies and applications are developed as solutions to complex systems from different fields.

### 2.2 Burgeoning interdisciplinary field

The division of disciplines allows each discipline to focus on a closely related set of variables in the real world from its unique perspective.[26] This lets them rapidly analyze and address problems of interest without becoming entangled in the complexity of actual systems. However, in the context of complex systems, variables that are usually separated across different disciplines also exhibit nonlinear connections. Thus, interdisciplinary approaches are needed to integrate knowledge and techniques from diverse disciplines, involving more nonlinear relationships (i.e., with squared terms or even higher powers) among a broader set of linked variables. Currently, no agreement exists on how to conduct interdisciplinary research. In the emerging interdisciplinary fields between social, computational, and complexity sciences, concepts such as social computing[27, 23] and computational social science[25, 7] have been proposed and witnessed substantial progress. We do not differentiate between these two concepts but rather discuss the two interdisciplinary paths from the perspectives of research methods and research paradigms, as illustrated in Figure. 1B.

On the one hand, advanced computational models and techniques have been continuously integrated into the analysis of social systems, resulting in numerous interdisciplinary methods that can effectively analyze complex social systems. With advancements in complex adaptive theory and computer simulation methods, social simulation methods have emerged as viable means of studying complex social phenomena.[28] Agent-based Modeling (ABM), a prominent social simulation method, employs a bottom-up approach to investigate emergent phenomena at the macro level resulting from interactions among heterogeneous agents at the micro level.[6] The development of ABM has facilitated the emergence of artificial societies,[29] which involve constructing digital society laboratories using ABM methods to evaluate policy effectiveness experimentally. Notable examples of artificial societies include SugarScape,[30] the artificial stock market,[31] and National Planning Scenario 1 (NPS1).[32] Besides social simulation, data mining methods focus on discovering knowledge about human dynamics,[33] poverty and wealth,[34] epidemics,[35] and other facets of social systems from large social datasets, offering new insights into social system analysis. Moreover, online experiment methods, enabled by internet platforms, provide an extensive pool of samples for social systems experiments.[8, 36, 37] This approach alleviates the three-horned dilemma of generality-control-realism[38] and the challenges of reproducing results[39] in social science. Along with the aforementioned three methods, there exist advanced techniques that facilitate the investigation of social interaction on a micro level, such as virtual reality[40] and hyperscanning.[41]

On the other hand, social and human factors are consistently integrated into the operation of various actual systems, providing fundamental research paradigms that support the blending of cyber, physical, and social spaces. Since the 1980s, with the continuous development of embedded systems, concepts such as ubiquitous sensing and pervasive sensing have emerged.[42] These ideas focus on embedding various computing devices within physical space, resulting in a close coupling between computing and physical resources through Cyber-Physical-Systems (CPS). With the growing prevalence of sensor-enabled smart devices and their tight connection to humans and society,[9] social space is gradually infiltrating CPS, driving its evolution toward more generalized socialization. Subsequently, various research paradigms have been developed to integrate social and human factors into CPS, including Cyber-Physical-Human Systems (CPHS),[10] Human-Cyber-Physical Systems (HCPS),[43] Social Cyber-Physical Systems (SCPS),[44] and Cyber-Physical-Social Systems (CPSS).[4] Among these, CPSS is the most prominent, deeply integrating Human, Machine, and IoT resources to enable coordination among cyber, physical, and social spaces.[14] Nowadays, CPSS has become a foundational paradigm for studying systems that encompass both social and engineering complexity in areas such as transportation,[45] agriculture,[17] healthcare,[46] and other fields,[47] involving activities such as sensing,[3, 48–50] computing,[51] producing,[15] operating,[52] and so on.

### 2.3 Parallel systems method with interdisciplinary nature

The increasing complexity of real-world systems poses numerous challenges, yet interdisciplinary studies have presented opportunities. In this context, the parallel systems method, inherently interdisciplinary, has emerged at the intersection of social and computational sciences, offering solutions to address the intricate social and engineering aspects of complex systems. The origin of the parallel systems method can be traced to the early 1990s when researchers were exploring open complex giant systems.[53] Its prototype was a "shadow system" designed for studying and evaluating a complex entity through its digital replicas.[54] In 2004, the ACP approach was introduced, marking the starting point of studying the parallel systems method, and it soon became the technical foundation for a 20-year journey in supporting the implementation of the parallel systems method.[11, 55, 56] The formalization of CPSS in 2010 fundamentally shaped the solution form of ACP-based parallel systems, enhancing its applications in various fields.[4, 57] As the research progressed, relevant studies appeared in a variety of fields, each interpreting and applying the parallel systems method in terms tailored to their unique perspectives. Researchers began to recognize that these studies need clear objectives and guidance. Therefore, in 2016, the concept of parallel intelligence[58] was proposed, providing a high-level encapsulation of the fundamental principles, guidelines, or methodologies of the parallel systems method.

When dealing with complex systems, the parallel system method adopts CPSS as its basic form to examine numerous elements in the system and construct artificial counterparts of these actual elements. Subsequently, computational experiments are conducted to explore complex interaction mechanisms between these elements and investigate potential behaviors of systems under different scenarios. Furthermore, the bidirectional interaction and coevolution of artificial and actual systems are facilitated through parallel execution. In applying the parallel systems method, it is imperative to assimilate domain knowledge from diverse disciplines to support activities such as modeling, computing, and executing. Supported by parallel intelligence, the ACP approach integrates various research methods such as simulation and data mining to form a more powerful intelligence for understanding complex systems. Parallel intelligence enables artificial systems to transform from system analyzers into data generators, overcoming challenges posed by modeling, experiments, and data defects.[12] Consequently, it enables the transition seamlessly from "small data" in the actual system to "big data" in artificial systems, and ultimately to "deep intelligence" in the solution by executing a continuous cycle of data, knowledge, and action between artificial and actual systems.[59]

To comprehensively investigate the current state of development in the parallel systems method, we conducted a snowballing literature review, yielding 215 relevant publications. The details of our search and screening procedures are delineated in the supplemental information. Our analysis reveals that research on the parallel systems method is currently focused on two main areas: parallel technologies and their applications. The former, as the method's advanced technical extension, will be presented in Section 4; the latter, as an interdisciplinary accomplishment of the method, will be discussed in Section 5.

## 3. PARALLEL SYSTEMS METHOD – AN INTERDISCIPLINARY SOLUTION

In this section, we probe into the details of the parallel systems method, as shown in Figure. 2. The parallel intelligence, as the objective of the method, aims to improve our understanding of complex systems and the ACP approach serves as the technical foundation to realize the objective.

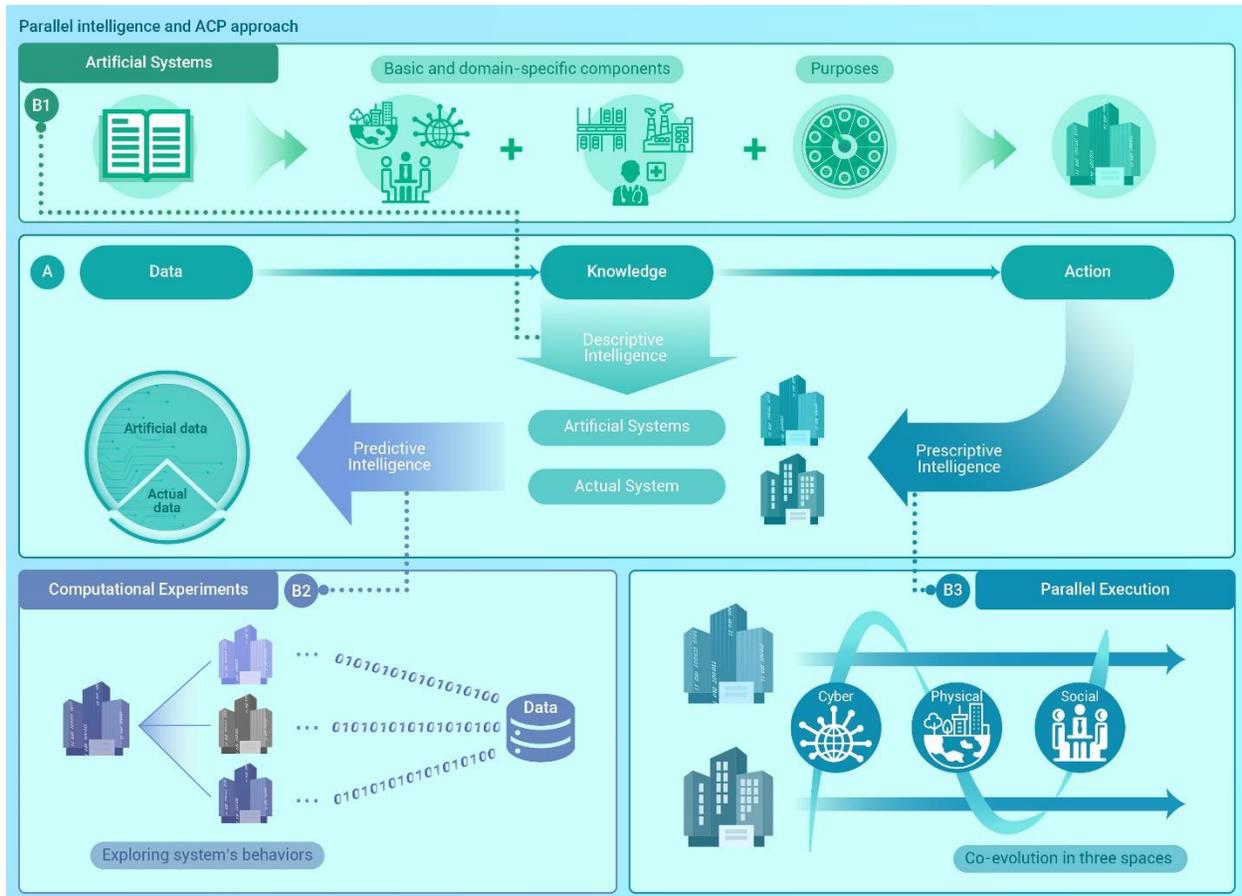

**Figure 2. Parallel systems method as an interdisciplinary solution.** (A)The objective of the parallel systems method is to achieve parallel intelligence, which involves a novel cycle of data, knowledge, and action between actual and artificial systems. This cycle is facilitated by descriptive, prescriptive, and predictive intelligence and the ACP approach provides the technical foundation for the implementation of three types of intelligence. (B1) Artificial systems are composed of basic and domain-specific components, designed to fulfill specific purposes. (B2) Computational experiments are conducted to generate data on the system's behaviors across different scenarios. (B3) Parallel execution is leveraged to facilitate the co-evolution of artificial and actual systems in cyber, physical, and social spaces.

### 3.1 Parallel intelligence – the objective

The long-standing cycle of data, knowledge, and action in scientific research is such that knowledge extracted from data informs action, leading to the acquisition of more desired data, which in turn refines the knowledge with new data. Wiener described this cycle as "circular causality", paving the path for the study of modern cybernetics and computational intelligence.[60] However, within the context of complex systems, this cycle faces challenges, primarily in the transition from action to data. Due to the nonlinear interactions and complex emergence mechanisms among the numerous elements in complex systems, obtaining desired data through actions performed in the actual system can be problematic. Moreover, economic, ethical, and moral considerations often make actions related to war or disasters difficult to execute. To mitigate challenges in the transition from action to data, simulation methods use artificial models as substitutes for the actual system within the cycle. However, owing to the inherent uncertainty of complex systems, a model deviation between the actual system and its artificial counterpart often exists.[59] This deviation leads to differences between the data distribution generated by simulation models and that of real-world data, ultimately resulting in cycle failure. Data mining methods focus on one phase of the cycle: the extraction of knowledge from large datasets. Though these datasets may be extensive, they can suffer from shortcomings such as incomplete or inaccessible information.[61] Moreover, data mining mainly depends on correlations within the data, and this reliance can pose difficulties in exploring causality. These constraints can affect the reliability and validity of the knowledge extracted.

Parallel intelligence aims to establish a novel cycle of data, knowledge, and action, integrating research methods such as simulation and data mining. It employs three distinct kinds of intelligence, —descriptive, predictive, and prescriptive—to enact this cycle. Akin to simulations, artificial systems can be constructed as digital counterparts of an actual system with the help of descriptive intelligence. Unlike a complete replication, parallel intelligence focuses on mirroring the actual system's behavior within its artificial constructs. This enables the artificial systems to generate data mimicking what the actual system might produce. Predictive intelligence explores potential behaviors of actual systems under different scenarios. By merging the generated big data with the actual system data, it forms comprehensive datasets. These datasets, in turn, facilitate the extraction or training of more reliable knowledge or intelligent models.[62] Serving as the final step, prescriptive intelligence aids in implementing actions based on the refreshed knowledge or intelligent models. It fosters the coevolution of artificial and actual systems toward desired outcomes, creating new data in the process.

Given the innate uncertainty and dynamism of complex systems, universal solutions are not applicable. Artificial systems, in the context of parallel intelligence, are not replacements for the actual system but function alongside it.[63] They must coexist within the same cycle with the actual system, and this cycle must be carried out simultaneously between them. This ensures that artificial systems can capture the dynamic characteristics of the actual system. Meanwhile, predictive intelligence should be engaged to probe as many possible behaviors of the actual system as possible, recognizing and navigating the inherent uncertainty of complex systems.

### 3.2 ACP approach – the technical foundation

The ACP approach comprises three components: artificial systems, computational experiments, and parallel execution. These represent the specific implementations of descriptive intelligence, predictive intelligence, and prescriptive intelligence, respectively.

*Artificial systems.* Artificial systems are usually constructed based on a diverse range of knowledge. By employing techniques such as knowledge automation[64] and knowledge representation,[65] software-defined objects, environments, and events are crafted as artificial components to mirror various elements in the actual system. These software-defined artificial components, which can be categorized into two groups: basic and domain-specific, are then integrated to form cohesive software-defined artificial systems.[66] The basic components encapsulate more than just the physical aspects of the actual systems; they also encompass cyber and social dimensions, including structures such as artificial platforms[57] and artificial populations.[67] Besides, domain-specific components vary cross various complex systems. For instance, artificial hospital systems include subcomponents such as artificial diagnosis, treatment, and nursing subsystems,[52] while artificial population systems comprise elements such as artificial cognition and human subsystems.[67] In the realm of agriculture, artificial systems are made up of components such as artificial genotypes, environments, crop growth, and farmer behavior subsystems.[68] Moreover, modeling methods manifest diversity across various fields. The ABM method has been widely employed to construct artificial systems, enabling each agent to engage in straightforward interactions. Complex network models are also widely used to characterize the complex relationships and interactions among agents in artificial systems.[69] In parallel vision research,[70] computer graphics and virtual reality (VR) technologies are employed for constructing artificial scenarios and simulating diverse elements present in real-world physical spaces such as illumination, weather conditions, and camera configurations. Learning models further contribute to modeling human cognitive systems.[71] Moreover, knowledge graphs[46] and mathematical models[72] are leveraged in artificial healthcare and artificial industry systems, respectively. The recent emergence of Large Language Models (LLM), including models such as ChatGPT, offers exciting possibilities for constructing generative agents with human behavior in artificial systems.[73]

Furthermore, a single actual system can correlate to several artificial systems, each serving different purposes. For instance, distinct artificial traffic systems can be constructed to represent different facets such as historical traffic conditions, normal and average performances, optimal and ideal operations, as well as worst-case scenarios for disaster and emergency management.[74]

*Computational experiments.* Computational experiments can be conducted in three distinct modes, each serving a specific purpose. In the learning and training modes, the data generated from experiments are integrated with actual scenarios or cases. This fusion facilitates knowledge acquisition and the accumulation of experience for operators and managers in real-world systems. In the experiment and evaluation mode, computational experiments are performed more extensively and intensely to analyze and predict the behavior of the actual system under various scenarios. In the control and management modes, computational experiments are conducted to optimize control schemes or management plans, with consideration given to the differences between actual and artificial systems.

Xue et al.[75] presented a range of fundamental techniques for conducting computational experiments, encompassing the construction, design, analysis, and validation of experimental systems. These computational experiments vary in form across different studies. For instance, parallel testing has employed semantic diagram models to generate challenging testing tasks that bolster the autonomous capabilities of vehicles.[76, 77] Learning models, such as Generative Adversarial Networks (GANs), have been integrated with computational experiments in parallel learning, resulting in the creation of traffic data,[78] realistic images of rare driving scenes,[79] and control schemes for traffic engineers.[80] Recently, scenarios engineering has emerged as a technique to produce large volumes of scenario data in parallel systems, enhancing the training of reliable and trustworthy foundation models.[81, 82]

*Parallel execution.* Parallel execution aims to facilitate collaborative and iterative optimization of both artificial and actual systems, utilizing insights from computational experiments. The techniques employed for achieving parallel execution differ among various spaces. In cyber space, parallel execution is commonly realized through the optimization of software, algorithms, and models, such as the various machine learning models used in parallel learning.[83] In the physical space, parallel execution can be achieved through the optimization of existing scheduling and control procedures in actual systems, such as real-time economic generation dispatching and control in power systems,[84] or dynamic scheduling and operation control in high-speed railway systems.[85] In social space, parallel execution becomes more intricate, with past research employing methods such as "information guidance" or "data-to-mind" methods[45] strategies. Examples include distributing energy coupons or energy-saving information to incentivize power users[86] and influencing pedestrian path selection and evacuation behavior through strategic signage.[87]

Furthermore, it is crucial to implement corrections and updates to artificial systems. There is no universal solution for managing or controlling complex systems, as new problems, needs, and trends continually emerge. These must be dynamically integrated into artificial systems in real-time. Consequently, optimized solutions can be derived from these artificial systems to steer the ongoing development and evolution of the actual system.

## 4. PARALLEL TECHNOLOGIES

Built on the ACP-based technical foundation, the parallel systems method is constantly enriched through integration with cutting-edge technologies from various disciplines, leading to the emergence of novel technologies such as parallel learning, parallel blockchain, and parallel cognition, as shown in Figure 3. These hybrid technologies enhance the capabilities of the parallel systems method, enabling it to tackle complex problems. Concurrently, the growth of parallel technologies assists in overcoming intrinsic challenges associated with these technologies. To ensure conciseness, this section will delve into five major parallel technologies, considering their overlapping and successive relationships (see supplemental information for details of the selection procedure).

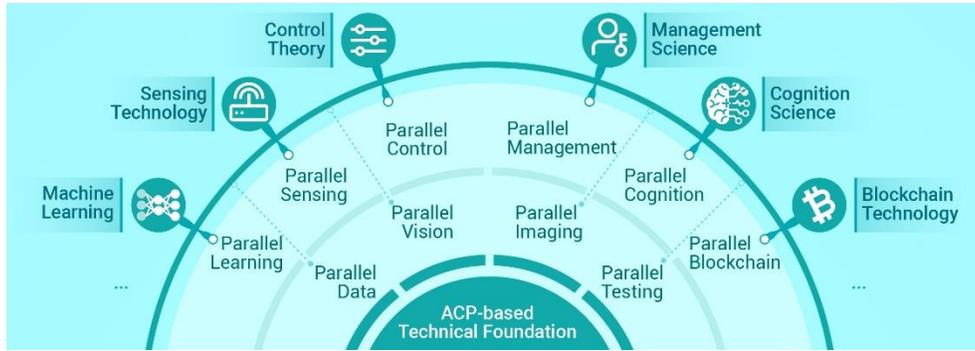

**Figure 3. Parallel technologies.** Various parallel technologies are proposed by integrating advanced technologies from different disciplines into the ACP-based technical foundation.

### 4.1 Parallel learning

Parallel learning, essentially a technical embodiment of parallel intelligence, has become a prominent parallel technology. Traditional machine learning methods grapple with issues such as inefficient exploration, poor generalization, and data scarcity, especially within the framework of complex systems.[62] Leveraging the ACP foundation, parallel learning enhances data augmentation, online knowledge updating, and efficient exploration of the action space. Such advancements enable the deployment of machine-learning-based solutions in analyzing and managing complex systems.[88, 89] In parallel learning, advanced models such as transfer learning, predictive learning, deep learning, reinforcement learning, and GAN are utilized to generate diverse synthetic datasets.[78] These datasets fused with actual system data, offer substantial training resources for the above models, culminating in more powerful and reliable intelligence, capable of providing precise and robust solutions for complex systems. Parallel learning found applications in areas such as traffic control,[90, 80, 91] traffic prediction,[92, 93] and energy systems management,[88, 89] further enhanced by parallel data.[94, 95]

Furthermore, parallel learning has been combined with parallel vision[70] and parallel imaging[96] to generate photorealistic images and scene data, addressing challenges related to visual perception and understanding. The advantages of synthetic data are twofold. First, it tackles the issue of imbalanced data distribution. For instance, in autonomous driving[97], current datasets mainly comprise normal driving scenarios while rare events like accidents are underrepresented, a phenomenon known as the long-tail effect.[98] Parallel learning techniques, such as SST-GAN[79] and LoTR,[99, 98] are capable of generating a diverse array of scarce data to compensate for this deficiency.[100, 101] Second, annotating large-scale data from real-world scenarios is cumbersome and labor-intensive, often resulting in a limited range and scale of labeled datasets.[102] Parallel learning provides a solution by generating ample labeled data for various tasks, including object positioning, motion trajectory tracking, semantic segmentation, depth estimation, optical flow analysis, and other vision-related labels.[70]

The efficacy of parallel learning has been empirically validated, but the disparity between artificial and real data should not be disregarded.[98] Ongoing research into Sim2Real may offer solutions to this issue.[103, 104] Besides, Miao et al.[105] have explored the interplay between parallel learning in the context of science for AI and AI for science, reinforcing our discussion on the interdisciplinary essence of the parallel systems method.

**4.2 Parallel blockchain**

Blockchain represents an innovative decentralized infrastructure and a distributed computing paradigm. However, it faces challenges such as security threats, block inflation, wasted computing resources, irrational competition, and other issues that hinder its progress. To mitigate these challenges, parallel blockchain[106] was introduced. This concept adds computational experiments and parallel decision-making functions to traditional blockchain operation, enabling the parallel interaction and coevolution of actual and artificial blockchains.[107] Using data encryption, timestamp distribution, distributed consensus, and economic incentives, parallel blockchain enables decentralized peer-to-peer transaction coordination and collaboration within a distributed system without the need for mutual trust between nodes. This provides a secure and trustworthy environment for the interaction of various elements within complex systems.[108]

Parallel blockchain enables the implementation of Decentralized Autonomous Organizations (DAO), a modern and effective organizational and managerial framework for navigating uncertain, diverse, and complex environments.[109, 110] DAO can serve as a hub for integrating cyber, physical, and social spaces. It allows for seamless storage and analysis of cyber data on blockchain; facilitates the digitalization and registration of physical devices, assets, and entities as smart properties on blockchain; and ensures privacy-protected, fair, and reliable human-machine and human-human interactions through smart contracts recorded on blockchain.[111, 112] Powered by parallel blockchain and DAO, numerous parallel applications have been developed in the fields of transportation,[113] healthcare,[114] and agriculture.[115]

### 4.3 Parallel control and management

In a broad perspective, the primary objective of many parallel technologies is to enhance management and control, as the ACP-based technical foundation essentially builds upon adaptive control methods in complex systems.[11] In a narrow perspective, parallel control can be regarded as a data-driven computational control approach in the control field.[116] For instance, Song et al.[117] applied parallel control to distributed parameter systems, exploring various cases such as the vibration control of Longmen crane, cooperative control of multiagent systems, chemical process control of plug flow reactor, and the control of flexible beam system. To address optimal control issues in discrete-time time-varying nonlinear systems, parallel control was further combined with adaptive dynamic programming and heuristic dynamic programming techniques.[118, 119] The self-learning-optimized parallel control method has been investigated to assess the theoretical performance of parallel control for the first time, laying a solid foundation for future research and expanding its application. This exploration also facilitates the advancement of interpretability in artificial intelligence.[12] Traditional organizations and management typically follow a top-down pyramid structure, which can restrict innovation potential and interaction efficiency.[110] Bolstered by parallel blockchain and knowledge automation, parallel management offers a decentralized and autonomous solution for management that highlights features of simple intelligence, provable security, flexible scalability, and ecological harmony.[120] Parallel management targets not only physical organizations but also artificial or computational counterparts, as well as their mechanism for virtual-real interactions.[121]

### 4.4 Parallel sensing

Sensing activities serve a pivotal function in bridging information transmission between artificial and actual systems. Traditional Sensing, restricted by hardware and software constraints of physical sensors, can be limiting. To counteract these constraints, Shen et al.[122] proposed parallel sensing that redefines sensors as a combination of actual physical sensors and virtual software-defined sensors, using the parallel systems method to enhance their functionality. In parallel sensing, physical sensors operate at discrete time intervals to conserve energy, while virtual sensors function continuously to provide compensatory data. This method has been applied to parallel vision,[70] parallel point clouds,[123] and parallel light fields.[124] Beyond enhancing sensor performance, the crowdsensing campaigns can also be optimized. Crowdsensing intelligence, a novel framework grounded in the parallel systems method, leverages the collective intelligence of heterogeneous sensing participants to gather data and information from CPSS.[48–50] A parallel vehicular crowdsensing system has also been developed to produce high-fidelity data for various traffic scenarios, sensing tasks, and road networks to guide the optimization of sensing campaigns.[57]

### 4.5 Parallel cognition

The investigation of parallel cognition represents a crucial stride toward exploring social dimensions through the ACP approach. Human cognition, with its complexity and uncertainty, has often been neglected or overly simplified in research. Cognitive science is conventionally interdisciplinary and primarily employs experimental, inductive, modeling, and validation paradigms. These may fall short in dealing with complex systems involving multiple individuals with substantial heterogeneity and dynamics. Moreover, ignoring human imperfections such as fatigue and faults can result in conflicts in decision-making between humans and machines, or even severe accidents. Parallel cognition[71] was formulated as a computational approach to interpret, learn, predict, and prescribe individual behaviors under specific conditions in given tasks. A hybrid learning method, founded on psychological models and user behavioral data, was proposed to adaptively understand individual cognitive knowledge. The validation of parallel cognition was conducted on two types of complex systems: social group decision-making in urban transportation and cognitive visual reasoning in human-in-the-loop systems. These studies demonstrate its effectiveness in fostering human-machine cooperation in complex engineering and social systems.[67] Although the current research on parallel cognition is relatively limited, its potential is evident. It offers a viable approach to delve deeply into social complexity and paves the way for the creation of more realistic and nuanced parallel systems.

## 5. PARALLEL APPLICATIONS

Leveraging technical foundations and parallel technologies, parallel applications in the basic form of CPSS have been successfully deployed across various domains such as transportation, healthcare, industry, agriculture, society, ecology,[125] education,[126] and artistic creation.[47, 127] We mainly exhibit them in five domains as illustrated in Figure. 4. Parallel applications aim to tackle the management and control challenges inherent in complex systems, while simultaneously enhancing their efficiency and sustainability. Additionally, they provide personalized solutions for individual needs such as transportation services and medical diagnosis and treatment.

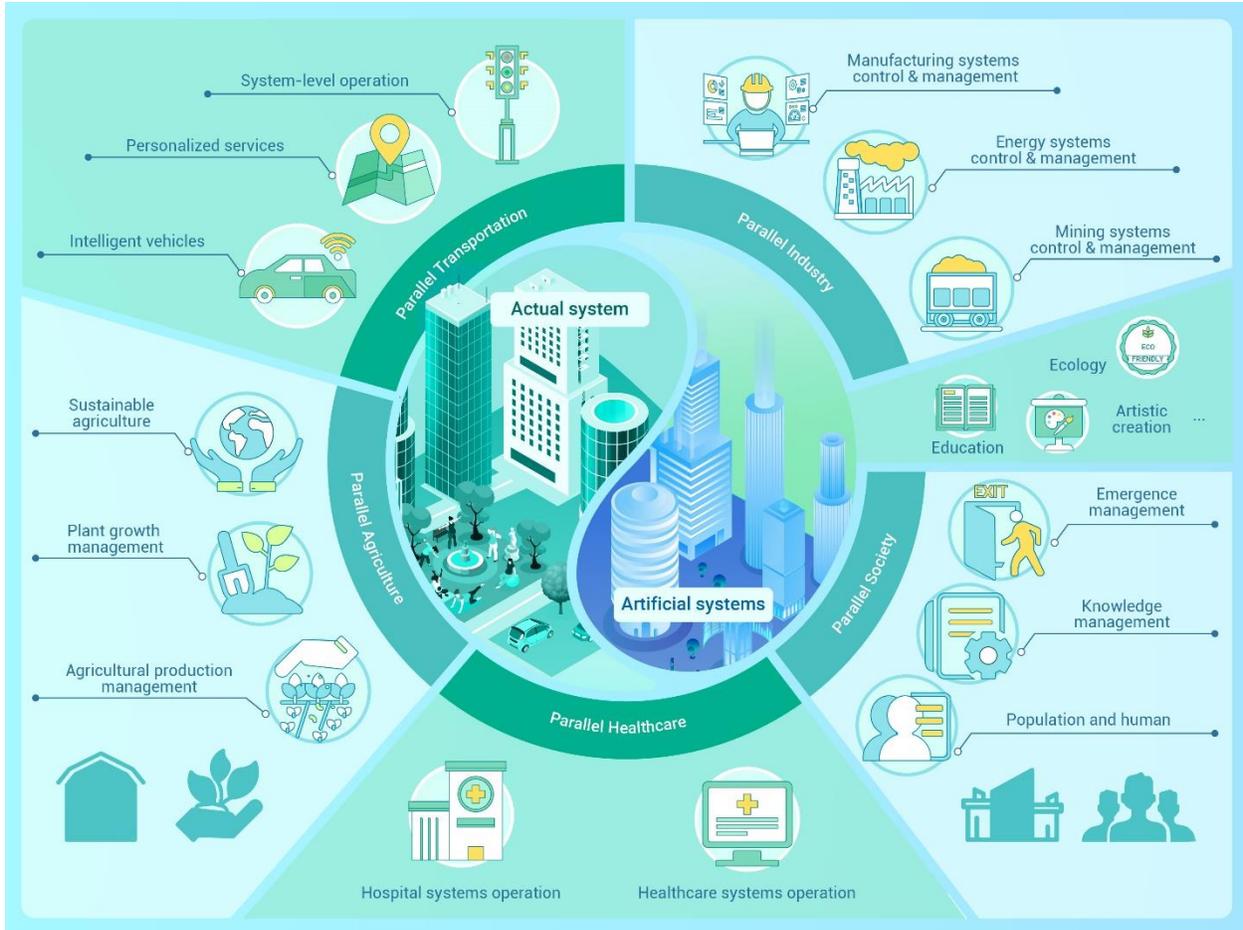

**Figure 4. Implementation and applications.** Various parallel applications have been developed to solve problems in complex systems from different domains.

### 5.1 Parallel transportation

As shown in Figure. 5, parallel transportation emerged as one of the earliest fields of parallel application. The swift growth of modern cities has given rise to complex issues in contemporary Intelligent Transportation Systems (ITS), including congestion, pollution, and accidents. These challenges require the development of system-level management and control schemes for ITS.[74] Integrating aspects such as traffic information collection, analysis, modeling, prediction, control management, and measurement feedback, Parallel transportation Management Systems (PtMS) offer a comprehensive solution. They are instrumental in evaluating ITS behavior under various scenarios,[45, 128–130] enhancing traffic prediction[78, 94] and control.[131–133] PtMS has been successfully deployed in cities such as Binzhou, Taicang, and Suzhou, China.[129] Sustainability within ITS has emerged as a pivotal focus,[134] leading to research on ACP-based energy-efficient schemes[135] and parallel emission regulatory frameworks.[136] Beyond system-level research, parallel transportation also extends personalized services, including parking guidance[137] and route planning.[19] Moreover, intelligent vehicles seen as potential solutions to intricate problems in ITS, benefit from parallel technologies, such as parallel learning,[79] parallel vision,[99] and parallel testing.[77, 76] This technological support has led to the development of applications such as parallel driving[138–142] and parallel vehicles[143, 144] to enhance interactions between intelligent vehicles, human drivers, and the environment,[145] as well as strengthen vehicle intelligence.[146, 147]

The field of parallel transportation has amassed numerous accomplishments, both in cutting-edge techniques and practical projects. Moving forward, in addition to technical challenges, legal and ethical considerations will need to be addressed to pave the way for real-world implementation of applications such as parallel driving and PtMS.

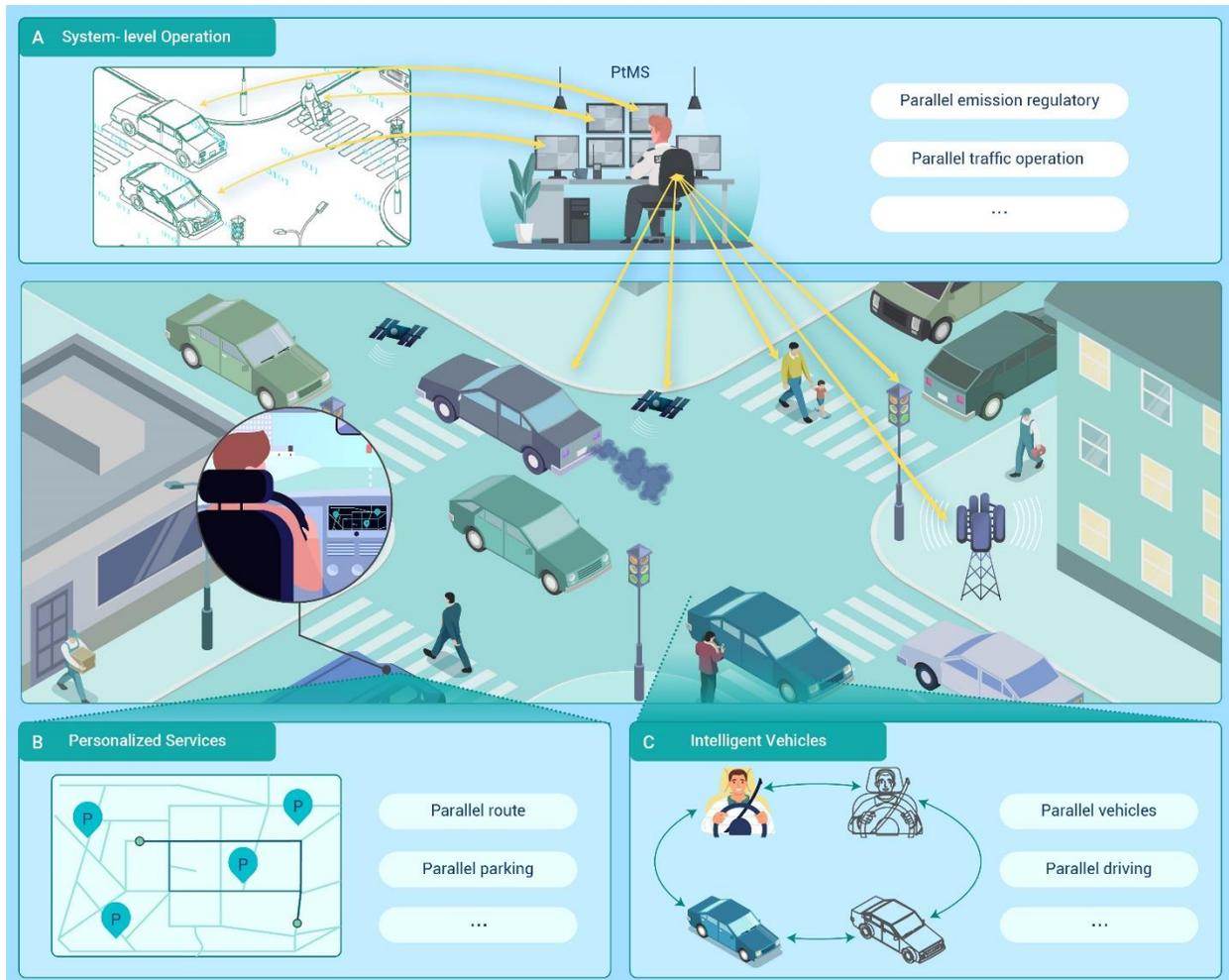

**Figure 5. Parallel applications in parallel transportation.** They are roughly divided into three related parts: a) system-level operations, such as emission regulatory and traffic operation; b) personalized services including routing planning and parking guidance; and c) intelligent vehicles designed for the implementation of autonomous driving.

**5.2 Parallel healthcare**

To address the social, humanistic, and scientific challenges in healthcare, a new infrastructure known as parallel healthcare has emerged based on the ACP approach.[52, 148] The two main areas of application lie in the operations of healthcare systems and hospital systems. Smart healthcare systems have played a crucial role in advancing modern medicine; however, their practical implementation still faces challenges, including inadequate cross-border collaboration among experts and insufficient personalized diagnoses and treatments. To address these issues, Wang et al.[114] proposed parallel healthcare systems that distill the synthesized data from both actual and artificial systems into precise knowledge or profound intelligence for individual-specific healthcare problems, using parallel technologies. Multiple parallel healthcare systems have been developed to collaborate with doctors in diverse fields, encompassing Gout,[149] surgery,[150, 151] hypertension,[46] ophthalmia,[152] and skin disease.[153]

Furthermore, the parallel systems method serves as a guiding framework for developing the next generation of intelligent hospital operation systems.[154] A parallel hospital can be regarded as a digital representation of a hospital, encompassing hospital operators, and all participants involved in medical activities.[52] This approach facilitates dynamic and rational scheduling or pre-allocation of hospital resources, thereby enhancing resource utilization and the quality of medical services. Supported by parallel blockchain and DAO, a parallel hospital can establish a federated ecology in the healthcare field by integrating resources from multiple medical institutions to improve overall system performance.[52] Furthermore, this research provides fundamental assurance for advancing remote healthcare and helps mitigates disparities in the allocation of medical resources.

Current research on parallel healthcare systems primarily relies on case studies and focuses on various specific diseases. Given the complexity of the human body, which consists of numerous systems, there is an imperative to develop comprehensive parallel healthcare systems in the future that encompass a wider range of physical and mental elements related to human health. To achieve a parallel hospital federation ecology, future studies must develop standards and norms to promote resource sharing among different institutions.

**5.3 Parallel industry**

Within the CPSS framework, the parallel industry integrates traditional industrial systems, such as manufacturing, mining, and energy systems, with elements in the social space, including social demands, worker status, and user behavior. This integration forges unprecedented opportunities for the effective management and control of industrial systems, simultaneously enhancing system reliability and operational efficiency, and promoting sustainable development. Parallel manufacturing[155] employs knowledge automation to analyze and distill social demands from social intelligence for product research and production planning, facilitating a swift response to market shifts.[15] Furthermore, parallel workers can supplant human workers in performing most physical and mental tasks, thus achieving cost-effective, efficient, and zero-inventory manufacturing.[16] The parallel systems method is employed to enhance the dependability and operational efficiency of energy systems, including nuclear power plants,[156] microgrids,[88] and the energy internet.[157] Additionally, parallel mining[158, 159] has been implemented and deployed at over 20 mining sites throughout China, significantly boosting production efficiency and safety standards in open-pit mines and fostering the mining industry's sustainable development.

Although the parallel industry has shown promising results in the areas aforementioned, particularly in parallel mining, the consideration of factors stemming from social space is still in its infancy and somewhat oversimplified. Further exploration must incorporate the intricacies of complex systems and delve into social and human factors more comprehensively.

**5.4 Parallel agriculture**

In light of the growing global population, agricultural systems face mounting pressures to produce enough food. Parallel agriculture aims to achieve sustainable agriculture with potential applications spanning all stages: pre-production tasks such as scheduling, market and demand analysis, and plant optimization; inter-production tasks such as planting management, environmental control, soil analysis, fertilization, spraying, irrigation, and the use of pesticides and herbicides; and post-production tasks including harvest, storage, processing, transportation, sales, and logistics scheduling.[68] Currently, the parallel systems method is leveraged to enable precision management of plant growth.[160] Kang et al.[17] proposed an agricultural CPSS for agricultural production management services based on the ACP approach. The system integrates social and physical sensors, including wholesale market prices of agricultural products and environmental data from daylight greenhouses, to offer decision-making support for planting plans through the ACP approach. This approach can improve economic efficiency while reducing labor and fertilizer wastage, thus promoting sustainable development.[161]

While the fundamental framework for parallel agriculture is in place, lingering challenges such as information asymmetry, data scarcity, and infrastructure inadequacy demand attention. Further case studies and empirical research are indispensable for propelling the successful implementation of parallel agriculture.

**5.5 Parallel society**

Empowered by advanced ICT, the swift spread of information and real-time interplay between online and offline environments has augmented the complexity of managing individuals and organizations within social systems.[23] Parallel societies utilize the ACP approach to establish society laboratories that offer decision-making support and wisdom for managing and controlling social systems. For instance, in social emergency management,[162] the parallel systems method has been employed to address public health crises[163, 164] as well as emergency evacuations[87, 165–168] across various scenarios. In human–machine hybrid CPSS, the parallel systems method has been employed to investigate human cognition, taking into account human heterogeneity and uncertainty through parallel humans and parallel populations.[71, 67] Prototypes of parallel humans and populations have been implemented for social security, social group decision-making, and advancing human-in-the-loop systems in complex engineering. Furthermore, parallel management has been applied to organize, manage, and measure knowledge work, reflecting the transition from manual labor to knowledge work in future intelligent societies.[120, 169]

Due to the intricate nature of social systems, a majority of the current research on parallel society is predominantly case-based. Moreover, the authenticity of these applications warrants further scrutiny. The future development of parallel society applications could potentially be facilitated through collaborative data-sharing mechanisms with governments and companies, enabling validation and verification in a data-driven manner.

## 6. CHALLENGES AND PERSPECTIVES

At present, the parallel systems method faces multifaceted challenges, including internal challenges related to its existing theory and technologies, external challenges posed by advanced Artificial Intelligence (AI), and persistent ethical and legal challenges. Meanwhile, we can anticipate the potential of this method to promote sustainable development and interdisciplinary communication and cooperation.

### 6.1 Challenges

The parallel systems method is likely to encounter three major challenges: internal challenges tied to the implementation of the ACP approach, external challenges driven by the development of AI, and ever-present ethical and legal challenges that always exist in human societies.

*Internal challenges.* Implementing the ACP approach presents internal challenges, as deviations in the model can grow with system complexity. *1) Hard to model.* For some complex systems, unclear internal mechanisms, incomplete knowledge systems, and limited data complicate the creation of corresponding artificial systems. While it is unnecessary for artificial systems to fully replicate the behavior of actual systems, there is still no consensus on determining the validity and equivalence of artificial systems.[170] *2) Hard to experiment.* In terms of computational experiments, many factors in complex systems have mutual effects and nonlinear interactions. These interactions may lead to causal inversion and spurious correlation, making it arduous to design a comprehensive experiment plan to explore the behaviors of complex systems. *3) Hard to control.* In terms of parallel execution, controlling actual system behaviors directly is often difficult due to the uncertainty of social factors and the free will of humans. In certain situations, behaviors can only be indirectly influenced, making parallel execution challenging to quantify, implement, and assess. Additionally, researchers must adapt the ACP approach to complex systems' diverse characteristics, a process where cognitive biases may inadvertently play a role, potentially causing negative influence.[67]

*External challenges.* The parallel systems method faces challenges from the fast-paced growth of AI. The rapid evolution of AI has spurred innovations across various disciplines.[171] Especially, the recent advancements in foundation models have achieved remarkable accomplishments in fields such as computer vision and natural language processing.[172] Navigating these advancements presents challenges for researchers across various fields, including those working with the parallel systems method. For instance, how can emerging AI technologies foster theoretical and technical innovation within the parallel systems realm? How will integrating new AI technologies with the parallel systems approach enhance problem-solving capabilities for complex systems, and how might this approach inspire breakthroughs in artificial general intelligence? We have already witnessed some preliminary studies on integrating foundation models with the parallel systems method,[82, 173] and the contribution of the parallel systems method to advancing reliable and trustworthy AI has been recognized.[81, 174]

The external challenges posed by AI development may evolve into opportunities for the parallel systems method in the future.

***Ethical and legal challenges.*** Ethical and legal challenges cannot be overlooked, and they mainly encompass three aspects. *1) Liability dispute.* In fields such as parallel society, parallel driving, and parallel healthcare, mistakes in decision-making, traffic accidents, and medical mishaps caused by machines or algorithms might lead to legal liability. *2) Privacy disclosure.* The implementation of the parallel systems method often necessitates engagement with social space and may require the collection of fine-grained individual data. Specifically, constructing artificial systems usually requires multi-modals and multi-source data from various aspects. This can create a significant risk of personal privacy disclosure through object association and matching within these data. *3) Human labor replacement.* Parallel applications in various domains enable efficient collaboration among humans, machines, and algorithms. However, they can also lead to reduced human resource utilization. For instance, the deployment of parallel manufacturing has cut the number of workers on a production line from 36 to just one.[15] This reduction might exacerbate societal concerns about new technologies supplanting human roles. Beyond these three aspects, there exist other ethical and legal challenges that necessitate attention, including algorithmic fairness and biases, intellectual property law, safety, and transparency.[175]

## 6.2 Perspectives

We anticipate that the parallel systems method will have far-reaching implications, offering critical insights in two aspects: promoting sustainable development and facilitating interdisciplinary research.

***Promoting sustainable development.*** The parallel systems method could become a cornerstone in the pursuit of sustainable development.[176] *1) Improve system efficiency.* By facilitating interactions between artificial and actual systems, this method paves the way for more effective cooperation among various elements, as well as more judicious resource allocation in complex systems. For instance, studies on parallel driving and parallel vehicles have demonstrated their potential in reducing greenhouse gas emissions and air pollution.[134] Meanwhile, parallel agriculture empowers precision agriculture, enabling meticulous resource planning and eco-friendly agricultural production.[68] Parallel mining enhances the efficiency and safety of open-pit mines while promoting environmentally responsible mining practices.[158] *2) Develop three types of humans.* To promote efficient cooperation between humans and other elements such as machines, algorithms, and advanced AI technologies, future parallel applications will introduce three classifications of humans: digital, robotic, and biological. Digital and robotic humans will take over the bulk of mental and physical labor, alleviating the intense demands placed on biological humans to manage complex systems. At present, the deployment of these three types of humans in fields such as autonomous driving,[142] knowledge management,[120] artistic creation,[127] and industrial production,[15, 16] has already demonstrated significant advancements in collaboration between humans and various other elements.

*Facilitating interdisciplinary communication and cooperation.* The parallel systems method has the potential to facilitate interdisciplinary communication and cooperation. *1) Serve as a consensus.* Different disciplines possess distinct perspectives and methodologies for addressing problems, necessitating a consensus among them to facilitate effective communication and cooperation. The parallel intelligence and ACP-based technical foundation can serve as the consensus to aid experts from different disciplines in gaining a comprehensive view of the entire problem-solving process for complex systems and comprehend their respective roles in the different stages. *2) Resolve conflict and inconsistency.* The implementation of the parallel systems method is anticipated to address conflicts between diverse disciplines and inconsistencies within individual fields that often arise from challenges in theory testing. For example, there is a controversy among scholars from diverse fields regarding the efficacy of ABM techniques;[170] in the field of social science, two theoretical models regarding social contagion and collective behavior have had a significant impact, however, they are logically incompatible.[177] The parallel systems method, being application-driven and tailored to specific complex systems, allows for practical validation of diverse models and methods. Through this validation, conflicts and inconsistencies can be reconciled based on practical outcomes. *3) Satisfy diverse disciplines.* The parallel systems method strives to advance knowledge across various disciplines. Situated in what has been referred to as "Pasteur's quadrant",[177, 178] it reveals that application-based research can also promote fundamental understanding within each discipline. Parallel applications offer solutions to the management and control of complex systems while satisfying epistemic values across different disciplines.[179] For example, the parallel systems method can yield explanatory insights (e.g., human dynamics and disease outbreaks[69, 164, 166]) that are typically concerned by social science, while it can also be used to make predictions (e.g., traffic flow and

accidents[18, 93, 94]) that are emphasized by computational science. *4) Become a mature field.* The parallel systems method has the potential to evolve into a comprehensive and independent interdisciplinary research direction. This could prompt a restructuring of academic curricula, allowing interdisciplinary researchers to cultivate a blend of skills and knowledge, underpinned by new thinking, philosophy, science, method, and technology.[60] The accumulated perspectives, philosophical concepts, and cultural aspects of interdisciplinary research through the parallel systems method will likely play a significant role in this evolution.

## 7. CONCLUSION

This review has delved into the origins and fundamental concepts of the parallel systems method, shedding light on its objectives and technical foundation. The exploration of advanced parallel technologies and applications serves as a testament to the interdisciplinary accomplishments of the method. Despite the challenges that still need to be addressed, the future of the parallel systems method looks bright. It is potential to contribute significantly to sustainable development and act as a unifying force across various disciplines, facilitating interdisciplinary communication and cooperation.

## ACKNOWLEDGMENTS

This work is supported by the National Natural Science Foundation of China (62173337, 62276272, 62202477, 21808181, 72071207, 72025405, 72088101), Special Key Project of Biosafety Technologies (2022YFC2604000) for the National Major Research & Development Program of China, Social Science Foundation of Shanghai (2022JG204-ZGL730), National Social Science Foundation of China (22ZDA102), and Hunan Science and Technology Plan Project (2020TP1013, 2020JJ4673, 2023JJ40685). This work is dedicated to the National University of Defense University (NUDT) for its 70$^{th}$ anniversary.

## AUTHOR CONTRIBUTIONS

Y. Z., Z.-Q. Z., S.-H. Q., and B. C. wrote and edited the manuscript. J.-C. H., X. L., and F.-Y. W. supervised and revised the manuscript. Other authors collected and collated the publications included in this review. All authors contributed to the article and approved the submitted version.

## DECLARATION OF INTERESTS

The authors declare no competing interests.

## SUPPLEMENTAL INFORMATION

It can be found online at XXXXX.

## LEAD CONTACT WEBSITE

The lead contact websites are https://sihangqiu.com/ (S.-H. Q.) and www.homexinlu.com (X. L.)